\documentclass[12pt]{article}     
\usepackage{amsfonts}                                                           \usepackage{amssymb}
\usepackage{graphics,psboxit,amsmath}
\usepackage{subfigure}
\usepackage{graphicx}

\def\hybrid{\topmargin -20pt    \oddsidemargin 0pt 
        \headheight 0pt \headsep 0pt
        \textwidth 6.25in       
        \textheight 9.5in       
        \marginparwidth .875in
        \parskip 5pt plus 1pt   \jot = 1.5ex}

\hybrid

\def\baselinestretch{1.2}

\catcode`\@=11

\def\marginnote#1{}

\def\draftlabel#1{{\@bsphack\if@filesw {\let\thepage\relax
   \xdef\@gtempa{\write\@auxout{\string
      \newlabel{#1}{{\@currentlabel}{\thepage}}}}}\@gtempa
   \if@nobreak \ifvmode\nobreak\fi\fi\fi\@esphack}
        \gdef\@eqnlabel{#1}}
\def\@eqnlabel{}
\def\@vacuum{}
\def\draftmarginnote#1{\marginpar{\raggedright\scriptsize\tt#1}}

\def\draft{\oddsidemargin -.5truein
        \def\@oddfoot{\sl preliminary draft \hfil
        \rm\thepage\hfil\sl\today\quad\militarytime}
        \let\@evenfoot\@oddfoot \overfullrule 3pt
        \let\label=\draftlabel
        \let\marginnote=\draftmarginnote
   \def\@eqnnum{(\theequation)\rlap{\kern\marginparsep\tt\@eqnlabel}%
\global\let\@eqnlabel\@vacuum}  }

\def\preprint{\twocolumn\sloppy\flushbottom\parindent 2em
        \leftmargini 2em\leftmarginv .5em\leftmarginvi .5em
        \oddsidemargin -.5in    \evensidemargin -.5in
        \columnsep .4in \footheight 0pt
        \textwidth 10.in        \topmargin  -.4in
        \headheight 12pt \topskip .4in
        \textheight 6.9in \footskip 0pt
        \def\@oddhead{\thepage\hfil\addtocounter{page}{1}\thepage}
        \let\@evenhead\@oddhead \def\@oddfoot{} \def\@evenfoot{} }

\def\numberbysection{\@addtoreset{equation}{section}
        \def\theequation{\thesection.\arabic{equation}}}

\def\underline#1{\relax\ifmmode\@@underline#1\else
        $\@@underline{\hbox{#1}}$\relax\fi}

\def\titlepage{\@restonecolfalse\if@twocolumn\@restonecoltrue\onecolumn
     \else \newpage \fi \thispagestyle{empty}\c@page\z@
        \def\thefootnote{\fnsymbol{footnote}} }

\def\endtitlepage{\if@restonecol\twocolumn \else \newpage \fi
        \def\thefootnote{\arabic{footnote}}
        \setcounter{footnote}{0}}  

\catcode`@=12
\relax

\def\figcap{\section*{Figure Captions\markboth
        {FIGURECAPTIONS}{FIGURECAPTIONS}}\list
        {Figure \arabic{enumi}:\hfill}{\settowidth\labelwidth{Figure
999:}
        \leftmargin\labelwidth
        \advance\leftmargin\labelsep\usecounter{enumi}}}
 \relax
\def\tablecap{\section*{Table Captions\markboth
        {TABLECAPTIONS}{TABLECAPTIONS}}\list
        {Table \arabic{enumi}:\hfill}{\settowidth\labelwidth{Table
999:}
        \leftmargin\labelwidth
        \advance\leftmargin\labelsep\usecounter{enumi}}}
 \relax
\def\reflist{\section*{References\markboth
        {REFLIST}{REFLIST}}\list
        {[\arabic{enumi}]\hfill}{\settowidth\labelwidth{[999]}
        \leftmargin\labelwidth
        \advance\leftmargin\labelsep\usecounter{enumi}}}
 \relax
\makeatletter
\newcounter{pubctr}
\def\publist{\@ifnextchar[{\@publist}{\@@publist}}
\def\@publist[#1]{\list
        {[\arabic{pubctr}]\hfill}{\settowidth\labelwidth{[999]}
        \leftmargin\labelwidth
        \advance\leftmargin\labelsep
        \@nmbrlisttrue\def\@listctr{pubctr}
        \setcounter{pubctr}{#1}\addtocounter{pubctr}{-1}}}
\def\@@publist{\list
        {[\arabic{pubctr}]\hfill}{\settowidth\labelwidth{[999]}
        \leftmargin\labelwidth
        \advance\leftmargin\labelsep
        \@nmbrlisttrue\def\@listctr{pubctr}}}
 \relax
\makeatother
\newskip\humongous \humongous=0pt plus 1000pt minus 1000pt

\newif\ifdtup

\relax

\usepackage[]{graphicx}
\usepackage{amsmath,latexsym,amssymb}
\usepackage{subfigure}
\usepackage{comment}

\DeclareMathOperator{\del}{\ensuremath{\partial}}
\DeclareMathOperator{\db}{\ensuremath{\bar \partial}}
\DeclareMathOperator{\di}{\ensuremath{\mathrm{d}}}
\DeclareMathOperator{\rank}{rank}

\newcommand{\h}{\ensuremath{h}}
\newcommand{\hb}{\ensuremath{\bar{h}}}

\newcommand{\F}[2]{{f^{#1}_{\phantom{#1}#2}}}

\newcommand{\Deriv}[2]{\frac{\di #1}{\di #2}}
\newcommand{\setN}{{\mathord{\mathbb N}}}
\def\braket#1{\mathinner{\langle{#1}\rangle}}
\def\braket#1{\mathinner{\langle{#1}\rangle}}
\renewcommand{\ss}[1]{\textsc{#1}}

\begin{document}

\begin{titlepage}
\begin{center}

\hfill hep--th/0512086\\
\hfill CPHT-RR034.0605\\
\hfill LPTENS-05/39\\

\vskip .5in

{\Large \bf Renormalization-group flows and charge transmutation in string
  theory\footnote{Partially supported by the EU under the contracts
    MEXT-CT-2003-509661, MRTN-CT-2004-005104 and MRTN-CT-2004-503369}}

\vskip 0.5in

{\bf Domenico Orlando$^{1,2}$},\phantom{x} {\bf P.Marios Petropoulos}$^1$ \phantom{x}
and  {\bf Konstadinos Sfetsos}$^{3,1}$
\vskip 0.1in

${}^1\!$
 Centre de Physique Th\'eorique de l'\'Ecole
  Polytechnique\footnote{Unit\'e mixte du CNRS et de l'\'Ecole
    Polytechnique, UMR 7644.}  91128 Palaiseau, France

\vskip .1in

${}^2\!$
Laboratoire de Physique Th\'eorique de l'\'Ecole Normale
  Sup\'erieure\footnote{Unit\'e mixte du CNRS et de l'\'Ecole Normale
    Sup\'erieure, UMR 8549.}  24 rue Lhomond, 75231 Paris Cedex 05,
  France

${}^3\!$
Department of Engineering Sciences, University of Patras,\\
 26110 Patras, Greece

\end{center}

\vskip .4in

\centerline{\bf Abstract}
\noindent
We analyze the behaviour of heterotic squashed-Wess--Zumino--Witten
  backgrounds under renormalization-group flow. The flows we consider
  are driven by perturbation creating extra gauge fluxes. We
  show how the conformal point acts as an attractor from both the
  target-space and world-sheet points of view. We also address the
  question of instabilities created by the presence of closed
  time-like curves in string backgrounds.


\end{titlepage}
\vfill
\eject

\def\baselinestretch{1.2}
\baselineskip 20 pt
\noindent

\tableofcontents

\section{Introduction}

The purpose of this note is to analyze the behavior of certain
string backgrounds under world-sheet renormalization-group flows.
In our set-up, these flows are driven by world-sheet operators,
which create, in the target space--time, extra $U(1)$, electric
or magnetic, gauge fields and push the string off criticality.

World-sheet renormalization group has been investigated both from
the general conformal field theory (\textsc{cft}) and from the geometrical,
target space viewpoints
\cite{Zamolodchikov:1986gt,Friedan:1980jm}. The motivations are
diverse: study the stability of the background against
off-critical excursions, search for new critical string
backgrounds, eventually explore string theory off-shell, etc.

In the presence of ``impurities" such as branes or orbifold fixed points in
non-compact target spaces, or when background electric or magnetic fields are
switched on, tachyons may in general
appear~\cite{Russo:2001tf,Kiritsis:1994ta}. World-sheet renormalization-group
techniques are then useful for investigating the relaxation process of the
original unstable vacuum, towards a new, stable infrared fixed point. Such
relaxation is usually a tachyon
condensation~\cite{David:2001vm,Harvey:2001wm,Freedman:2005wx}, that can be
accompanied by emission of particles (in the form \textit{e.g.}  of charged
pairs)~\cite{Bachas:1992bh, Pioline:2005pf}.

The presence of closed time-like curves can also trigger decays.
It was argued years ago~\cite{Hawking:1991nk} that
gravitational solutions with such chronological pathologies might
naturally evolve towards chronologically safe backgrounds. This
has been recast more recently in the framework of string
vacua~\cite{Israel:2003cx,Costa:2005ej} with
some preliminary results. It is clear that one would gain
insight by studying renormalization-group flows in an
appropriately chosen parameter space for families of string
backgrounds.

World-sheet renormalization group can be studied directly at the level
of the two-dimensional \textsc{cft}. Any relevant operator
can be used to leave the conformal point, and the necessary tools are
in principle available for computing the beta-functions and
determining the flows. This procedure is usually perturbative. It turns out that
it is trustful \cite{Zamolodchikov:1986gt}
in determining the new conformal point in the IR
only when the operators
responsible for the flow are marginally relevant (conformal dimensions
$\Delta=\bar \Delta =1$, but only at first order in the deformation parameter),
or almost relevant (conformal dimensions $\Delta=\bar \Delta =1 - \varepsilon$).
In the case of deformations with irrelevant operators we return back to the conformal
point towards the IR.
In the framework of Wess--Zumino--Witten (\textsc{wzw}) models \cite{Witten:1983ar} (which capture
\textit{e.g.} the $S^3$ and AdS$_3$ spaces with NS background fluxes),
such operators exist only at large level $k$.  Hence, their operator
product expansions involve a plethora of fields, and the actual
computation of their beta function is very intricate.
In order to overcome this difficulty, we will here use an alternative
method, more geometric and based on target-space techniques.

This note is organized as follows. In section 2 we
present a quick review of the heterotic squashed \textsc{wzw} models
\cite{Israel:2004vv,Israel:2004cd}. Then in section 3 we introduce a
perturbation and study the
system as the \textsc{rg} flow, in the corresponding two-dimensional $\sigma$-model,
takes it back to the conformal point.
In section 4 we show that this is consistent with the \textsc{cft} results.

%

\section{Squashed \textsc{wzw} models}
\label{sec:conformal-model}

One of the most appealing properties of \textsc{wzw} models is
that they allow for both an exact \textsc{cft} bidimensional
description and a simple spacetime interpretation in terms of
group manifolds. Current-current deformations allow to explore
their moduli space, leading in general to models that keep the
integrability properties but may lack a nice spacetime
description. Special attention is deserved by the asymmetric
deformations in which the two currents come from different sectors
of the theory; in this case, in fact, together with the nice
\textsc{cft} properties, the spacetime geometry remains simple to
describe in terms of squashed groups.

To be more concrete consider a heterotic \textsc{swzw} model on a
group $G$ of dimension $d$ and rank $r$. The asymmetric
current-current deformation is realized by adding the operator
\begin{equation}
  \mathcal{O} = \frac{\sqrt{kk_g}}{2\pi} \int \di^2 z \  \sum_{a,b=1}^r
  c_{ab} \left( J^{a} (z) - \frac{i}{k} \F{a}{\textsc{mn}}
    \psi^{\textsc{m}} \psi^{\textsc{n}} \right) \bar J^b (\bar{z} ) ,
\end{equation}
where $J^a$ are currents in the Cartan torus $T \subset G$,
$\psi^{\textsc{m}}$ are the fermionic superpartners and $\bar J^a$ are
anti-holomorphic currents belonging to the gauge sector. The
engineering dimension of the operator is obviously $\left( 1, 1
\right)$ and, as it has been shown in~\cite{Chaudhuri:1989qb},
$\mathcal{O}$ is truly marginal (\emph{i.e.} at every order in
deformation) for any value of the parameter matrix $c_{ab}$, since the
currents commute. In other words, with the aid of $\mathcal{O}$, we
reach an $r$-dimensional space of exact \textsc{cft}'s.

As described in~\cite{Israel:2004cd}, the background fields corresponding to
the new sigma--model can be read using a technique bearing many resemblances
to a Kaluza--Klein reduction\footnote{It would be a genuine reduction if we
  had done the construction in type \textsc{ii} or in a bosonic theory.
  In this case the current $\bar J^a$ would just be the anti-holomorphic
  derivative of an internal coordinate $X^a$.} and consist in a metric, a
Kalb--Ramond field and a $U(1)^r$ (chromo-)magnetic field. As announced above
the description remains simple and the all-order exact expression can be given in
terms of Maurer--Cartan currents $J^{\textsc{m}}$ on $G$ as follows:$\phantom{\h}$
\begin{equation}
  \label{eq:squashed-WZW}
  \begin{cases}
    \di s^2 = \displaystyle{\sum_{\mu \in G/T}} J^\mu J^\mu + \left( 1 - {\h}^2 \right)
    \displaystyle{\sum_{a \in T}} J^a J^a ,\\
    H_{[3]} = \frac{1}{2} f_{\mu \nu\rho} J^\mu \wedge J^\nu \wedge
    J^\rho & \mu \in G/T, \\
    F^a = \h \sqrt{\frac{k}{k_g}} \F{a}{\mu \nu} J^\mu \wedge J^\nu & \text{$\mu \in G/T$, $a \in T$},
  \end{cases}
\end{equation}
where we chose $c_{ab} = \h \delta_{ab}$. In particular we see that
the metric is the one of a squashed group \emph{i.e.} we still have
the structure of a $T$ fibration over $G/T$ but the radius of the
fiber changes with $\h$. A special value of the deformation parameter
is singled out: for $\h < 1 $ the metric is positive definite, while
for $\h > 1 $ the signature changes. The apparently singular $\h = 1 $
value can nevertheless be reached by a limiting procedure whose
geometrical interpretation is the trivialization of the fiber. We end
up with an exact \textsc{cft} on a $G/T$ background sustained by a
(chromo-)magnetic field.

The simplest example is given by $G = SU (2)$ where we have (in
Euler coordinates) the following background fields:
\begin{equation}
\label{eq:squashed-SU2}
  \begin{cases}
    \di s^2 = \di \theta^2 + \di \psi^2 + \di \phi^2 + \cos \theta \di
    \psi \di \phi - \h^2 \left( \di \psi + \cos \theta \di \phi
    \right)^2\ , \\
    B = \cos \theta \di \psi \wedge \di \phi\ ,\\
    A = 2 \h \left( \di \psi + \cos \theta \di \phi \right)\ ,
  \end{cases}
\end{equation}
corresponding, in the $\h \to 1$ limit, to a $S^2$ geometry.


\section{RG-flows for compact groups: geometric approach}
\label{sec:compact-groups}

We present here the geometric, target-space techniques for
analyzing \textsc{rg} flows in two-dimensional theories. These
techniques apply to any compact group. We will however expand on
the case of $SU(2)$ since it captures all the relevant features.

\subsection{The parameter space}
\label{sec:model}

The model that we have presented in the previous section is
conformal; for this reason we expect to find it as a fixed point
in an \textsc{rg} flow. To verify this claim let us introduce a
two-parameter family of $\sigma$~models generalizing the exact
backgrounds of Eq.~(\ref{eq:squashed-WZW}); a possible choice
consists in adding a new magnetic field, this time coming from a
higher dimensional right sector. Explicitly
\begin{equation}
  \label{eq:general-deformed}
  \begin{cases}
    \di s^2 = \displaystyle{\sum_{\mu \in G/T}} J^\mu J^\mu + \left( 1 - \h^2 \right)
    \displaystyle{\sum_{a \in T}} J^a J^a ,\\
    H_{[3]} = \frac{\hb}{2\h} f_{\mu \nu\rho} J^\mu \land J^\nu \land
    J^\rho & \mu \in G/T, \\
    F^a = \frac{\h + \hb}{2} \sqrt{\frac{k}{k_g}} \F{a}{\mu \nu} J^\mu \land
    J^\nu & \text{$\mu \in G/T$, $a \in T$}, \\
    \bar F^a = \frac{\h - \hb}{2} \sqrt{\frac{k}{k_g}} \F{a}{\mu \nu} J^\mu \land J^\nu & \text{$\mu \in G/T$, $a \in T$}
  \end{cases}
\end{equation}
and in particular for $SU(2)$:
\begin{equation}
\label{eq:2par-su2}
  \begin{cases}
    \di s^2 = \di \theta^2 + \di \psi^2 + \di \phi^2
+ \cos \theta \di \psi \di \phi - \h^2 \left( \di \psi + \cos \theta \di \phi \right)^2\ ,\\
    B = \frac{\hb}{\h} \cos \theta \di \psi \land \di \phi\ ,\\
    A = \left( \h + \hb \right) \left( \di \psi + \cos \theta \di \phi \right)\ ,\\
    \bar A = \left( \h - \hb \right) \left( \di \psi + \cos \theta \di \phi \right)\ ,
  \end{cases}
\end{equation}
where $\hb $ is a new parameter, describing the deviation from the
conformal point. It is clear that the above background reduces to the
one in Eq.~(\ref{eq:squashed-SU2}) in the $\hb \to \h $ limit.  In
particular we see that the metric is unchanged, the Kalb--Ramond field
has a different normalization and a new field $\bar A$ appears.  This
configuration can be described in a different way: the geometry of a
squashed sphere supports two covariantly constant magnetic
fields with charge $Q = \h + \hb$ and $\bar Q =\h - \hb$;
the \textsc{rg} flow will describe the evolution of these
two charges from a generic $\left( Q, \bar Q \right)$ to $\left( 2 \h,
  0 \right)$, while the sum $Q + \bar Q = 2 \h$ remains
constant.
 In this
sense the phenomenon can be interpreted as a charge transmutation of
$\bar Q $ into $Q$.
The conservation of the total charge is in fact a
consequence of having chosen a perturbation that keeps the metric
and only changes the antisymmetric part of the background.

We can also see the background in Eq.\eqref{eq:general-deformed} from
a higher dimensional perspective where only the metric and the
Kalb-Ramond field are switched on. Pictorially:
\begin{align}
  \label{eq:4d-2par-su2}
  g = \left(
    \begin{tabular}{ccc|c}
      & & & \\
      & $g_{\textsc{wzw}}$ & & $\h J_a$  \\
      & & &  \\ \hline
      & $\h J_a$ & & 1
    \end{tabular} \right) &&
  B = \left(
    \begin{tabular}{ccc|c}
      & & & \\
      & $\frac{\hb}{\h} B_{\textsc{wzw}}$ & & $\hb J_a$  \\
      & & &  \\ \hline
      & $-\hb J_a$ & & 0
    \end{tabular} \right)
\end{align}
where $g_{\textsc{wzw}}$ and $B_{\textsc{wzw}}$ are the usual metric
and Kalb--Ramond fields for the \textsc{wzw} model on the group $G$.
More explicitly in the $SU(2)$ case:
\begin{align}
  g = \begin{pmatrix}
    1 & 0 & 0 & 0 \\
    0 & 1 & \cos \theta & \h \\
    0 & \cos \theta & 1 & \h \cos \theta \\
    0 & \h & \h \cos \theta & 1
  \end{pmatrix} &&
  B = \begin{pmatrix}
    0 & 0 & 0 & 0 \\
    0 & 0 & \frac{\hb}{\h} \cos \theta & \hb \\
    0 & -\frac{\hb}{\h} \cos \theta & 0 & \hb \cos \theta \\
    0 & -\hb & -\hb \cos \theta & 0
  \end{pmatrix}
\end{align}
where the fourth entry represents the bosonized internal current.  In
particular this clarifies the stated right-sector origin for the new
gauge field $\bar A$. This higher dimensional formalism is the one we
will use in the following \textsc{rg} analysis.

\subsection{The renormalization group flow}
\label{sec:renormalization}

The $\sigma$-model in Eq.~(\ref{eq:4d-2par-su2}) is not conformal
for generic values of the parameters $\h $ and $\hb$; this is why
it makes sense to study its behaviour under the \textsc{rg} flow.
Following a dimensional-regularization scheme (see
\cite{Osborn:1989bu,Alvarez-Gaume:1981hn,Friedan:1980jm} and for
various applications
\cite{Balog:1996im,Balog:1998br,Sfetsos:1998kr,Sfetsos:1999zm}) we
consider the action
\begin{equation}
  S = \frac{1}{2\lambda} \int \di^2\! z
\left( g_{\textsc{mn}} + B_{\textsc{mn}} \right) \del X^{\textsc{m}} \db X^{\textsc{n}} ,
\end{equation}
where $g$ and $B$ are the fields in Eq.~(\ref{eq:4d-2par-su2}).  The
beta-equations at two-loop order in the expansion in powers of the overall
coupling constant $l$ and the field redefinitions for
the internal coordinates $X^i$ turn out to be:
\begin{equation}
  \begin{cases}
    \beta_{\lambda^\ast} =
 \Deriv{\lambda^\ast}{t} = - \frac{\lambda^{\ast 2}}{4 \pi} \left( 1-
      \frac{\hb^2}{\h^2} \right) \left( 1+ \frac{\lambda^\ast}{8 \pi}
\left( 1 - 3 \frac{\hb^2}{\h^2} \right) \right), \\
    \beta_{\h} = \Deriv{\h}{t} = \frac{\lambda^\ast \h}{8 \pi} \left( 1- \h^2 \right)
    \left( 1- \frac{\hb^2}{\h^2} \right)
\left( 1+ \frac{\lambda^\ast}{8 \pi} \left( 1 - 3 \frac{\hb^2}{\h^2} \right) \right) , \\
    \beta_{\hb} = \Deriv{\hb}{t} = - \frac{\lambda^\ast \hb}{8 \pi} \left( 1 + \h^2
    \right) \left( 1- \frac{\hb^2}{\h^2} \right)
\left( 1+ \frac{\lambda^\ast}{8 \pi} \left( 1 - 3 \frac{\hb^2}{\h^2} \right) \right) , \\
    X^i = X^i - \frac{\lambda^\ast}{16}
\left( 1 - \h^2 \right) \left( 1- 4 \frac{\hb^2}{\h^2} + 3 \frac{\hb^4}{\h^4} \right) ,
  \end{cases}
\end{equation}
where $\lambda^\ast = \lambda g^\ast$, $g^\ast$ being the dual Coxeter
number, is the effective coupling constant ($\lambda^\ast = N \lambda
$ for $G = SU(N)$). The contributions at one- and two-loop order are
clearly separated. In the following we will concentrate on the
one-loop part and we will comment on the two-loop result later.
Let us then consider the system:
\begin{equation}
\label{eq:compact-beta}
  \begin{cases}
    \beta_{\lambda^\ast} = \Deriv{\lambda^\ast}{t} = - \frac{\lambda^{\ast 2}}{4 \pi} \left( 1-
      \frac{\hb^2}{\h^2} \right) , \\
    \beta_{\h} = \Deriv{\h}{t} = \frac{\lambda^\ast \h}{8 \pi} \left( 1- \h^2 \right)
    \left( 1- \frac{\hb^2}{\h^2} \right), \\
    \beta_{\hb} = \Deriv{\hb}{t} = - \frac{\lambda^\ast \hb}{8 \pi} \left( 1 + \h^2
    \right) \left( 1- \frac{\hb^2}{\h^2} \right)\ .
  \end{cases}
\end{equation}
This can be integrated by introducing the parameter $z = \hb / \h$ which makes one
of the equations redundant. The other two become:
\begin{equation}
\label{eq:compact-red-system}
  \begin{cases}
    \dot \lambda^\ast = - \frac{\lambda^{\ast 2}}{4 \pi} ( 1 - z^2 ) ,\\
    \dot z = - \frac{\lambda^\ast z}{4 \pi} ( 1 - z^2 )\ .
  \end{cases}
\end{equation}
By inspection one easily sees that $\dot \lambda / \lambda = \dot z /
z $, implying $ \lambda (t) = C z (t)$, where $C$ is a constant.
This was to be expected since $C$ is proportional to the normalization
of the topological \textsc{wz} term. Since we are dealing with a
compact group it turns out that $C$ is, as in \cite{Witten:1983ar},
quantized with:
\begin{equation}
  C_k = \frac{2 \pi}{k},\ \ k \in \setN \ .
\end{equation}
Now it's immediate to separate the system and find that $z (t)$ is
defined as the solution to the implicit equation:
\begin{equation}
  - \frac{t}{2 k} = \frac{1}{z_0} - \frac{1}{z(t)} + \log \left[ \frac{
      \left( z(t) + 1 \right) \left( z_0 - 1 \right) }{\left( z(t) - 1
      \right) \left( z_0 + 1 \right)} \right]
\end{equation}
with the initial condition $z(0) = z_0$. A similar expression was found in
\cite{Braaten:1985is,Witten:1983ar}. The reason for this is, as pointed out
previously~\cite{Kiritsis:1995iu}, that the conformal model ($\hb = \h$) in its
higher-dimensional representation (the one in Eq.~(\ref{eq:4d-2par-su2}))
coincides with a $G \times H$ \textsc{wzw} model after a suitable local
field redefinition.

\begin{figure}
  \subfigure[$(z, \lambda) $ plane]{\includegraphics[width=7cm]{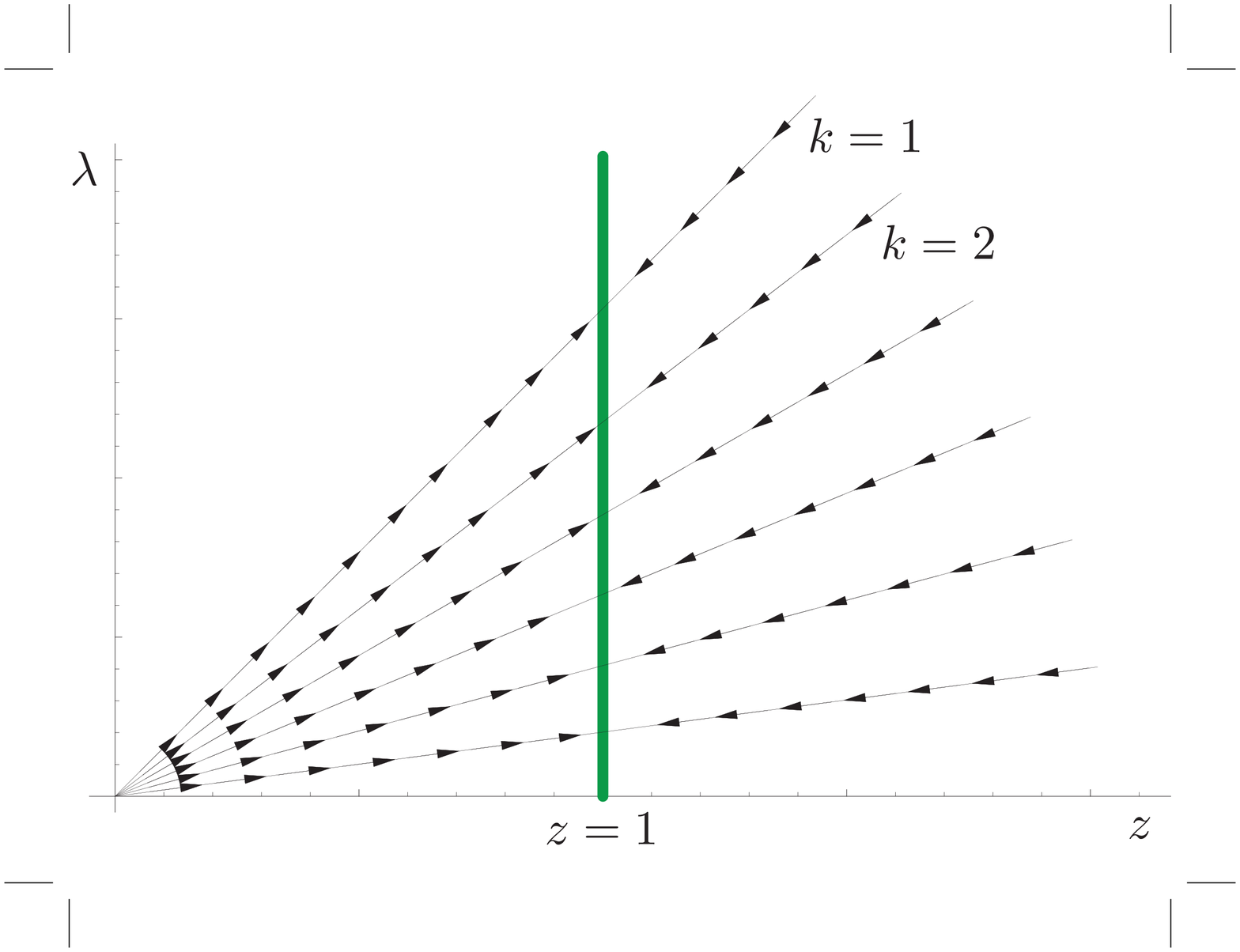}}
  \subfigure[$(\h, \hb)$ plane]{\includegraphics[width=7cm]{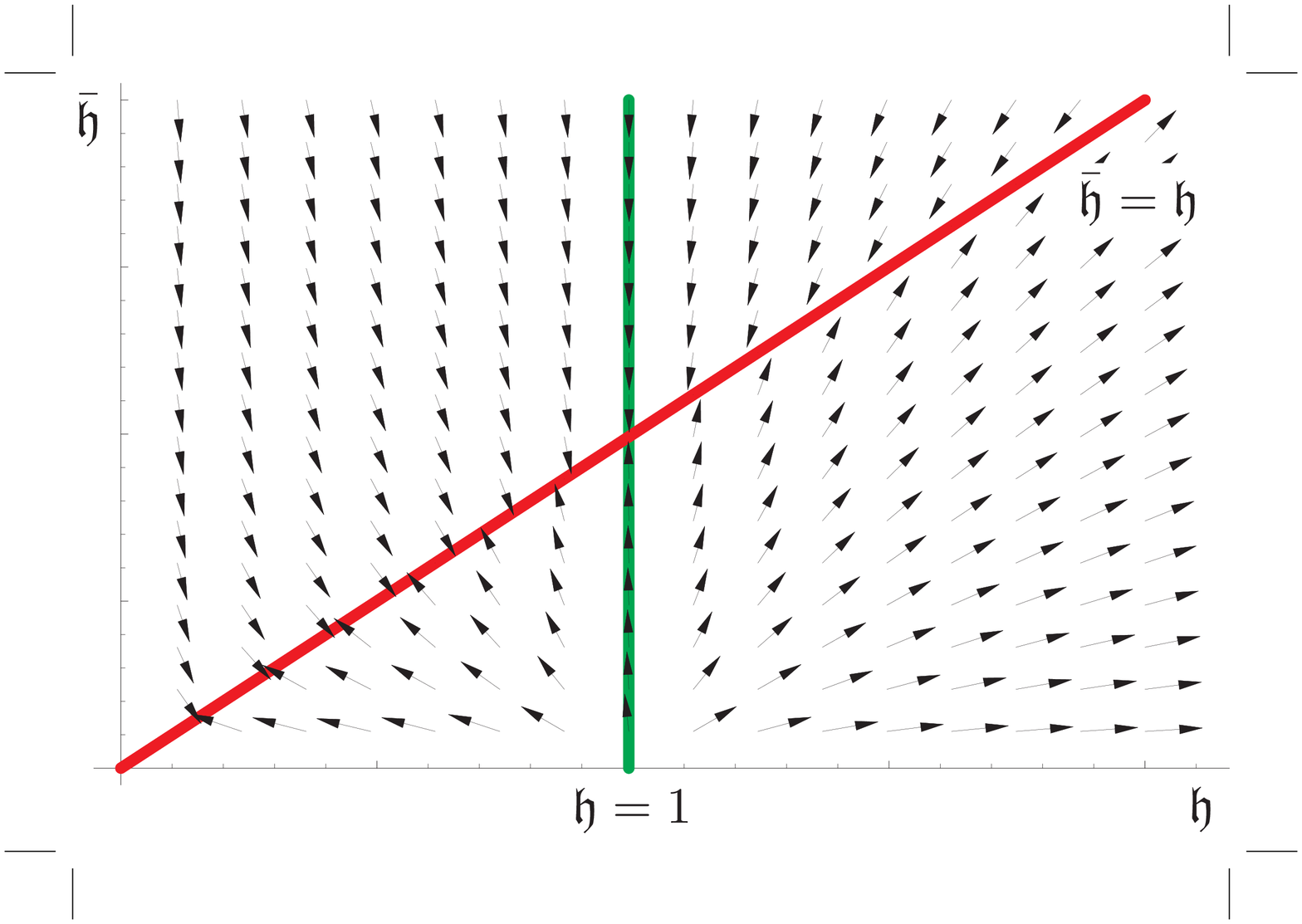}}
  \caption{Flow lines for the deformed (non-conformal) squashed
    \textsc{wzw} model in (a) the $(z, \lambda )$ and (b) the $(\h, \hb)$
    planes. The arrows point in the negative $t$ direction, \emph{i.e.}
    towards the infrared; in (a) we see how the squashed \textsc{wzw} model
    $z=1$ appears as an \textsc{ir} fixed point, in (b) how perturbing the
    conformal $\hb = \h $ model by increasing $\hb$ leads to a a new fixed
    point corresponding to a value of $\h$ closer to $1$.}
  \label{fig:flow-lines-su2}
\end{figure}

As it is usually the case in the study of non-linear dynamics, a
better understanding of the solution is obtained by drawing the
\textsc{rg} flow. In a $\left( z, \lambda^\ast \right)$ plane, the
trajectories are straight lines through the origin and only a discrete
set of them are allowed.  Moreover the line $z = 1 $ is an \textsc{ir}
fixed-point locus. This situation is sketched in
Fig.~\ref{fig:flow-lines-su2}(a). Just as expected the $z = \hb / \h = 1
$ point, corresponding to the initial exact model described in
Sec.~\ref{sec:conformal-model}, is an \textsc{ir} fixed point for the
\textsc{rg} flow.

Further insights can  be gained if
we substitute  the condition $\lambda^\ast  = C_k  \hb /  \h$ into the
system (\ref{eq:compact-beta}) thus getting:
\begin{equation}
  \label{eq:compact-HHb}
  \begin{cases}
    \Deriv{\h}{t} =  \frac{\hb}{4 k} \left( 1- \h^2 \right)
    \left( 1 - \frac{\hb^2}{\h^2} \right) , \\
    \Deriv{\hb}{t} = - \frac{\hb^2}{4 k \h} \left( 1 + \h^2 \right)
    \left( 1 - \frac{\hb^2}{\h^2} \right) .
  \end{cases}
\end{equation}
The flow diagram for this system in the $\left( \h, \hb \right)$
plane, Fig.~\ref{fig:flow-lines-su2}(b), shows how the system relaxes
to equilibrium after a perturbation. In particular we can see how
increasing $\hb$ leads to a a new fixed point corresponding to a value
of $\h$ closer to $1$.

We would like to pause for a moment and put the above results in
perspective. The target-space of the sigma-model under
consideration is a squashed three-sphere with two different
magnetic fields. Along the flow, a transmutation of the two
magnetic charges occurs: the system is driven to a point where one
of the magnetic charges vanishes. This fixed point is an ordinary
squashed-\textsc{wzw} (of the type studied in Sec.
\ref{sec:conformal-model}), that supports a single magnetic
charge.

As we pointed out in Sec. \ref{sec:conformal-model}, in the
squashed-\textsc{wzw}, the magnetic field is bounded by a critical
value, $\h = 1$. As long as $\h \leq 1$, the geometry is a genuine
squashed three-sphere. For $\h > 1$, the signature becomes Lorentzian
and the geometry exhibits closed time-like curves.  Although of
limited physical interest, such a background can be used as a
laboratory for investigating the fate of chronological pathologies
along the lines described above.  In particular we see that under the
perturbation we are considering the model shows a symmetry between the
$\h >1 $ and $\h<1$ regions. In fact the presence of closed time-like
curves doesn't seem to make any difference, but for the fact that
regions with different signatures are disconnected, \emph{i.e.} the
signature of the metric is preserved under the \textsc{rg} flow.
It is clear that these results are preliminary. To get a more reliable
picture for closed time-like curves, one should repeat the above
analysis in a wider parameter space, where other \textsc{rg} motions
might appear and deliver a more refined stability landscape.



\section{RG-flows for compact groups: CFT approach}
\label{sec:cft}

In order to make contact with genuine \textsc{cft} techniques, we must
identify the relevant operators which are responsible for the $(\h,
\hb)$ deformation of the $G \times H$ original \textsc{wzw} model ($H
= U(1)^{\rank G}$). At lowest approximation, all we need is their
conformal dimensions in the unperturbed theory.

Following~\cite{Zamolodchikov:1986gt},
let ${\cal L}_0$ be the unperturbed (conformal) action and $ \mathcal{O}_i$
the operators of conformal dimension $\Delta_i$. Consider the perturbed model,
with Lagrangian
\begin{equation}
  {\cal L} = {\cal L}_0 + g^i \mathcal{O}_i\ .
\end{equation}
The tree-level beta-functions read:
\begin{equation}
  \label{eq:cft-beta}
  \beta^i (g) = (\Delta_i-1)g^i\ ,
\end{equation}
where $g^i$ is supposed to be small, for the perturbative
expansion of $\beta^i$ to hold\footnote{One should be very careful in the
  choice of signs in these formulae. In~\cite{Zamolodchikov:1986gt} the time
  variable, in fact, describes the evolution of the system towards the infrared
  and as such it is opposite with respect to the $t = \log \mu $ convention
  that we used in the previous section (as in \cite{Witten:1983ar}).}.

The $G \times H$ primary operator we need can be written as follows:
\begin{equation}
  \mathcal{O} = \sum_{\ss{a,b}} \braket{ t^{\ss{a}} g t^{\ss{b}} g^{-1}}
\braket{ t^{\ss{a}} \del g g^{-1}  } \braket{ t^{\ss{b}} g^{-1} \db g}
=  \sum_{\ss{a,b}} \Phi^{\ss{ab}}  J^{\ss{a}} \bar J^{\ss{b}}\ ,
\end{equation}
where $\Phi^{\ss{ab}}$ is a primary field transforming in the adjoint
representation of the left and right groups $G$. As such, the total conformal
dimensions are \cite{Knizhnik:1984nr}
\begin{equation}
  \Delta = \bar \Delta = 1 + \frac{g^\ast}{g^\ast + k}\ ,
\end{equation}
where $g^\ast$ is the dual Coxeter number and as such the operator is
irrelevant (in the infrared).

Specializing this general construction to our case we find that the
action for the fields in Eq.~(\ref{eq:4d-2par-su2}) is:
\begin{equation}
\label{eq:cft-action}
  {\cal L} = \frac{k}{4 \pi} \left\{ {\cal L}_0 + \left( \frac{\h}{\hb}- 1\right)
\sum_{\ss{a,b}} \Phi^{\ss{ab}} J^{\ss{a}} \bar J^{\ss{b}}
+ \frac{\h}{\hb} \left( \h + \hb \right) \sum_i J^{a_i} \bar J^i
+  \frac{\h}{\hb} \left( \h - \hb \right) \sum_{i,\ss{a}} J^i \Phi^{a_i \ss{a}}
\bar J^{\ss{a}}  \right\} \ ,
\end{equation}
where $\ss{a}$ runs over all currents, $i$ over the internal currents
in $H$ and $J^{a_i}$ is the \textsc{wzw} current of the Cartan
subalgebra of $G$ coupled to the internal $\bar J^i$.  The extra terms
can be interpreted as combinations of operators in the $G \times
H$ model. The beta-functions are thus computed following
Eq.~(\ref{eq:cft-beta}) with the coupling $g=h/\hb-1$. We obtain
\begin{equation} \label{eq:hhb-beta}
 \Deriv{}{t}  \left( \frac{\h}{\hb} -1 \right)
    \Bigg|_{\hb = \h} = \frac{ g^\ast}{g^\ast + k }\left({h\over\hb}-1\right) + \cdots  =
\left(\frac{g^\ast}{k } - \frac{g^{\ast 2}}{k^2} \right)
\left({h\over\hb}-1\right) + \cdots\ ,
\end{equation}
where the dots after the first equality denote higher order terms
in the $(h/\hb-1)$-expansion and after the second equality, in
addition to that, higher order terms in the $1/k$-expansion.
This result is the same as the one in \cite{Knizhnik:1984nr} since, as we have mentioned,
there is the a local field redefinition that maps this model at the conformal point to
the $G\times H$ \textsc{wzw} model.
The above result
is to be compared with the results following from
Eq.~(\ref{eq:hhb-beta}) when they are expanded around
$\h=\hb$. We obtain:
\begin{equation} \label{eq:hhb-beta1}
 \Deriv{}{t} \left( \frac{\h}{\hb} -1 \right)
    \Bigg|_{\hb = \h} = \left(\frac{g^\ast}{k } -
    \frac{g^{\ast 2}}{k^2}\right)\left({h\over \hb}-1\right)+
{1\over 2}\left(- {g^\ast\over k} + 7 {{g^\ast}^2\over k^2}\right)\left({h\over \hb}-1\right)^2
+\cdots\ .
\end{equation}
We see that these results agree to first order in the coupling $h/\hb-1$.

The extra information that we obtain from this calculation is about
the interpretation for the two-loop beta-function we described
in the previous section. The one-loop
corrections to~(\ref{eq:cft-beta}) are of the form $ C_{ijk} \,
g^i \, g^j$, where $ C_{ijk}$ are related to the three-point function
of the unperturbed theory \cite{Zamolodchikov:1986gt}.
This coefficient is a measure of the
dimension of the operator $\mathcal{O}_i$ in the theory perturbed by
the set of all operators. Such a computation goes beyond
the scope of the present note. Nevertheless,
(\ref{eq:hhb-beta1}) predicts the coefficient of the term $(h/\hb-1)^2$ to second
order in the $1/k$-expansion and it seems that such a computation is feasible from
the \textsc{cft} viewpoint at least as a series expansion in $1/k$.

\section{Conclusions}
\label{sec:outro}

In this work, we have analyzed the phase space of squashed
\textsc{wzw} models, away from the original conformal point. Our
analysis is given in detail for the compact group $SU(2)$ and can be
generalized to any compact group. We have restricted ourselves to
deviations from the conformal point, generated by switching on
simultaneously two distinct magnetic (or electric) fields. The
corresponding backgrounds may have interesting interpretation in terms
of NS5-branes. We have investigated the phase diagram using
geometrical, target-space techniques, as well as standard \textsc{cft}
renormalization-group methods. Our results can be summarized as
follows: the squashed-\textsc{wzw} models are found, as expected, as
\textsc{ir} fixed points in the \textsc{rg} flow, and this result is
confirmed from both a target space and world-sheet point of view. The
field theory interpretation of this flow consists in what we have
called charge transmutation. One $U(1)$ charge transforms into another
$U(1)$ while the total charge is conserved. For large values of the
parameter $\h$ the backgrounds under consideration contain closed
time-like curves. These do not seem to change the behaviour of the
flow and the model remains stable, at least under the deformation we
consider.

This charge transmutation enters the class of phenomena that are
expected to take place when a metastable string background jumps to a
stable one through a non-critical path. These include tachyon
condensation, particle production and other interesting physical
phenomena: the \textsc{rg} flow around the conformal point is a tool
to get information on the dynamics of the relaxation.  Our geometrical
tools are well-fitted to describe the latter provided we allow for
more parameters in the phase space. A generalization of our approach
may also allow to address more thoroughly the issue of instabilities
triggered by the presence of closed time-like curves.  Of course this
is all very preliminary and in particular much still remains to be
done in clarifying the link between energy minimization and time
evolution in non-compact and time-dependent backgrounds. A first step
in this direction consists in investigating non-compact groups, like
$SL(2,\mathbb{R})$, for which some aspects (\emph{e.g.} related to
Zamolodchikov's C-theorem) of the underlying theory remain obscure.


\newpage
\vskip .4in

\centerline{ \bf Acknowledgments}

\noindent
We would like to thank C.~de Calan, E.~Kiritsis, C.~Kounnas,
N.~Toumbas, J.~Troost and especially V.~Dotsenko for illuminating
discussions.\\
P.M.~P.  and D.~O.  acknowlegde Patras University for
kind hospitality as well partial financial support by the INTAS
contract 03-51-6346 {\it Strings, branes and higher-spin gauge fields} and the EU under the
contracts MEXT-CT-2003-509661, MRTN-CT-2004-005104 and
MRTN-CT-2004-503369 and by the Agence Nationale pour la  Recherche, France,
contract 
05-BLAN-0079-01.\\
K.~S. acknowledges partial financial support by
the CNRS, the EU under the contract MRTN-CT-2004-005104, the INTAS
contract 03-51-6346 {\it Strings, branes and higher-spin gauge fields}, as
well as the Greek Ministry of Education programs $\rm \Pi Y\Theta
A\Gamma OPA\Sigma$ with contract 89194 and $\rm E\Pi AN$ with
code-number B.545.\\
Based on a talk given by D.~Orlando at the RTN meeting ``Constituents,
Fundamental Forces and Symmetries of the Universe'' in Corfu, Greece,
20--26 September 2005.

\bibliography{Biblia}

\end{document}